# GPU PaaS Computation Model in Aneka Cloud Computing Environments

*Shashikant Ilager[1], Rajeev Wankar[2], Raghavendra Kune[3] and Rajkumar Buyya[1]*

[1]**Clou**d Computing and **D**istributed **S**ystems (CLOUDS) Laboratory
School of Computing and Information Systems
The University of Melbourne, Australia
[2]School of Computer and Information Sciences
University of Hyderabad, Hyderabad, India
[3] Advanced Data Processing Research Institute
Department of Space, Secunderabad, India

**Abstract**

Due to the surge in the volume of data generated and rapid advancement in Artificial Intelligence (AI) techniques like machine learning and deep learning, the existing traditional computing models have become inadequate to process an enormous volume of data and the complex application logic for extracting intrinsic information. Computing accelerators such as Graphics processing units (GPUs) have become de facto SIMD computing system for many big data and machine learning applications. On the other hand, the traditional computing model has gradually switched from conventional ownership-based computing to subscription-based cloud computing model. However, the lack of programming models and frameworks to develop cloud-native applications in a seamless manner to utilize both CPU and GPU resources in the cloud has become a bottleneck for rapid application development. To support this application demand for simultaneous heterogeneous resource usage, programming models and new frameworks are needed to manage the underlying resources effectively. Aneka is emerged as a popular PaaS computing model for the development of Cloud applications using multiple programming models like Thread, Task, and MapReduce in a single container .NET platform. Since, Aneka addresses MIMD application development that uses CPU based resources and GPU programming like CUDA is designed for SIMD application development, here, the chapter discusses GPU PaaS computing model for Aneka Clouds for rapid cloud application development for .NET platforms. The popular opensource GPU libraries are utilized and integrated it into the existing Aneka task programming model. The scheduling policies are extended that automatically identify GPU machines and schedule respective tasks accordingly. A case study on image processing is discussed to demonstrate the system, which has been built using PaaS Aneka SDKs and CUDA library.

1. **Introduction**

The cloud computing has revolutionized the computing paradigm in recent years. It provides on-demand convenient access to a shared pool of configurable computing resources through the Internet [1]. Platform as a Service (PaaS) is a cloud service delivery model that provides tools for rapid development and deployment of applications in the cloud. Presently, most PaaS frameworks target to build cloud applications that use only CPU computing resources. Aneka [2] is one such PaaS framework, which helps to perform the aforementioned functionalities. Aneka provides APIs to build the distributed cloud applications using Task, Thread and Map Reduce programming models. The application can be deployed over private, public and hybrid clouds.

Due to recent advancements and breakthroughs in AI technology, many of the software applications are using these advanced AI techniques to boost their performance and the quality of their decision making in real time. Most of these applications have at least one or more module that makes use of machine learning algorithms or such available techniques for problem-solving. Meanwhile, GPUs have become the standard computing platform for machine learning algorithms due to its abundant computing cores which yield high throughput. GPUs usage has been driven by multiple use cases like autonomous vehicles, scientific computing, finance, healthcare and computational biology domains [3]. In addition, real-time applications like online gaming, video streaming, and Blockchain applications such as bitcoin mining demand high-end GPUs for their performance boost. To that end, applications have to be designed to use both the CPU and GPU resources together to cater the diverse user needs. The specific part of applications that are computationally intensive is delegated to the GPU to leverage the massive parallel capabilities of GPU while the CPU handles other imperative functionalities of the applications.

However, having GPU accelerators on every machine of the physical/ virtual cluster would increase cost and energy usage during processing [4]. Moreover, the throughput of cluster reduces since the majority of their GPUs are in the idle state for most of the time. One way to address this issue is by acquiring a minimum number of GPU instances and sharing the GPU workload across machines. This can be achieved through techniques like remote access and scheduling mechanisms.

Public cloud service providers like Amazon AWS[1] and Microsoft Azure[2] offer GPU instances for the end users. However, these offerings do not have platform support for distributed GPU computing that uses both CPU and GPU resources. The GPU clusters are managed (scheduling, provisioning, etc.) by either using platforms specific schedulers like Apache Spark or by using standard cluster resource management systems like SLURM or TORQUE (Linux based). Nevertheless, these platforms are application specific and are not desirable for many generic applications and managing these platforms on the cloud is an onus on users. In addition, most of the frameworks are targeted for Linux environment only which cannot be used for heterogeneous resource usage applications that are developed and deployed in a Windows operating system environment.

---

[1] https://aws.amazon.com/

To overcome the aforementioned limitations, the PaaS frameworks are needed which automatically manage the underlying resources with different scheduling policies in a seamless manner. Developers can leverage the standard API's provided by the PaaS platform to develop and deploy the application rapidly. Therefore, to achieve this, the extension should be made to the Aneka PaaS scheduling interfaces and task programming model to support the GPU computing, specifically, CUDA programming model. In the proposed model in this chapter, applications are built using .NET-supported programming languages targeting windows operating system environment. The advantages of these frameworks are including 1) Cost reduction 2) Provides transparent APIs to access the remote GPUs 3) Enables distributed GPU programming by extending the existing Aneka scheduler for GPU aware resources.

The problem that is addressed in this chapter can be stated as "How to effectively provide programming support for building the cloud applications that can use GPUs in the cluster where only some of the nodes in the cluster have GPU units". To solve this, a framework is required that supports resource management, including provisioning, monitoring, and scheduling services integrated. To that end, the Aneka provides all these integrated services to manage the CPU-based resources, and thus current Aneka architecture is extended to support for GPU resources as well.

The rest of the chapter is organized as follows: Section 2 describes the background topics that are relevant to this context such as Cloud computing, GPU computing, and Aneka. Section 3 briefly details the motivation for this study. Section 4 explains the related work as well as current practices in public clouds regarding GPU computing. Section 5 presents a proposed methodology to incorporate GPU model in Aneka and show the extended Aneka architecture and scheduling mechanism. The feasibility of the system is demonstrated in Section 6 with an image processing application for edge detection along with performance results that are carried out on a desktop-based private cluster. The open challenges and future directions are explained in Section 7, and finally, the chapter is summarized in Section 8.

---

[2] https://azure.microsoft.com

## 2. Background

This section briefly discusses Cloud computing, GPU computing, and concisely explain Aneka and its architectural elements.

### *2.1.   Cloud Computing*

Cloud computing offers access to shared pool of dynamically reconfigurable computing resources as on-demand pay-as-you-go computing basis. The virtualization technology in the cloud has enabled much-needed elasticity and allows to dynamically (de)provision the resources. The vast amount of such advantages has attracted many of the enterprises to move their IT infrastructure to the cloud. Besides, it offers a wide range of services at the different level of abstractions based on the user requirements. Accordingly, cloud services are broadly categorized into three types as follows:

1. **Infrastructure as a Service (IaaS):** This service model provides compute, network and storage services for end users. Most of the resources are virtualized and they can be scaled quickly both horizontally or vertically. Examples include Amazon AWS EC2, Microsoft Azure, and Google Cloud.

2. **Platform as a Service (PaaS):** This service model provides frameworks, tools, and SDKs to manage resources, build and deploy the applications rapidly. Users are relieved from procuring and managing the computing infrastructure. Examples include Google App Engine, Aneka.

3. **Software as a Service (SaaS):** This service model provides direct access to the software system. Users are relieved from both managing infrastructure and developing and maintaining the software applications. Examples include Gmail, Facebook, and Microsoft Office 365.

### 2.2. *GPU Computing*

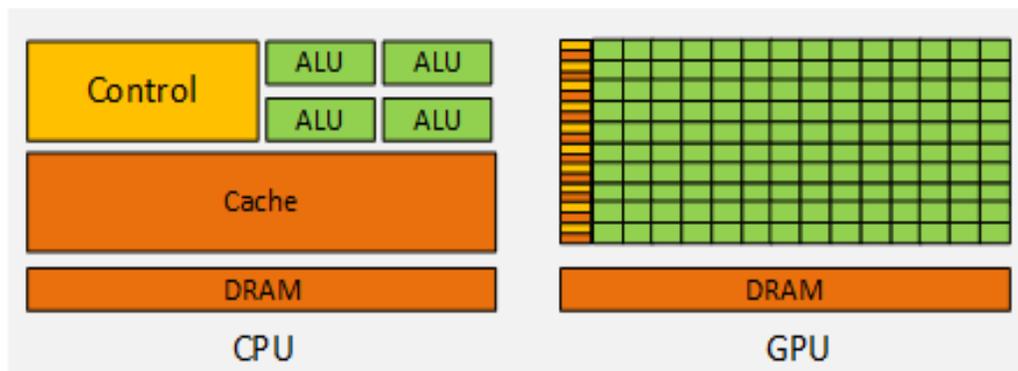

Figure 1. The Architecture of CPU vs. GPU

Graphical Processing Units (GPUs) have been extensively used as a fixed-function processor, built around the graphics pipeline, for processing images or videos [5]. With the advent of technology breakthroughs, GPUs have broadened their horizon and are also now used for general purpose computation and also called as General Purpose Graphic Processing Unit (GPGPU) [5].

The GPUs follow Single Instruction Multiple Data stream (SIMD) architecture and are efficient in processing parallel computing applications. They are connected to CPU though high-speed PCIe bus and are also known as co-processors or accelerators. In addition, a GPU follow

7multi-core architectures with low clock speed focused on high throughput, a contrast to CPU which focuses on low latency instruction execution with sophisticated branch prediction techniques. Each GPU core is less powerful than CPU core, however, the number of cores in GPU is higher than CPU which makes it as computing powerhouse compared to the latter.

GPUs have evolved in its architecture to support the general-purpose computation that now includes new memory hierarchy that has DDR memory, shared memory, cache memory, and registers which are essential elements in reprogrammable computing devices. With this, the user can arbitrarily read and write to memory which was not possible in older GPUs. As shown in Figure 1, CPUs have a limited number of cores and large cache memory, whereas GPU has a large number of cores with smaller caches which are suitable for parallel applications.

To easily develop the GPGPU applications, new programming models have been introduced. CUDA [6] is a one such GPU programming model developed by NVIDIA, it provides simple to use APIs with multiple programming languages support. OpenCL [7] is another opensource library targeted for multiple GPGPU vendors agnostic architectures.

### 2.3. *Aneka: Cloud Application Platform*

Aneka [2] is a complete Platform-as-a-Service framework supporting multiple programming models for the rapid development of applications and their deployment on distributed heterogeneous computing resources.

Aneka provides a rich set of .NET based APIs to developers for exploiting resources transparently that are available in cloud infrastructure and expressing the business logic of applications by using the preferred programming abstractions. Moreover, system administrators can leverage on a collection of tools to monitor and control the deployed infrastructure. The infrastructure can be built upon a public cloud available to anyone through the Internet, or a private cloud constituted by a set of nodes with restricted access. More importantly, the Aneka can also be used to set up a hybrid cloud that includes computing nodes from both private and public cloud. It supports multiple programming models and developers can choose a suitable model to build the cloud-native applications according to the application needs. The four programming models that are supported by the Aneka are:

- **Task model:** Applications are built with a set of independent bag-of-tasks;
- **Thread model:** Applications are composed with a set of distributed threads;
- **MapReduce model:** Applications that demand a large amount of data processing and follows MapReduce [10] programming model; and
- **Parameter sweep model:** Applications that are designed for execution of the same task over different ranges of values and datasets from a given parameter set.

The Aneka framework has been designed based on Service Oriented Architecture (SOA). Services are the basic elements of Aneka platform that allows to incorporate new functionalities or replace the existing one by overriding the current implementation. The abstract description of these services are as follows: *Scheduling* - the job of scheduling is service is to map the tasks to the available resources. *Provisioning* - this service can be used to acquire the resources (computing elements in terms of virtual or physical machines).

The network architecture of Aneka follows the master and worker nodes; the former orchestrates and manages all the resources viz. monitoring, scheduling, pricing, etc while the latter acts as computing elements which are responsible for executing the jobs assigned by its master.

Different services of Aneka are managed with layered architecture. Fabric layer of Aneka provides automated services like high availability, resource provisioning, hardware profiling, etc.

Foundation layer provides services like billing, storage, resource reservation and license and policing while Application service layer provides a set of programming models. The detailed architecture of Aneka is shown in  Figure 4  discussed in Section 5.2

## 3. Motivation

The important aspect of any computing infrastructure is to attain high resource utilization and reduce operational cost, it is obvious in the case of the economically driven clouds. To achieve high resource utilization, resources must be carefully shared among multiple applications or users. The following are the benefits of providing GPGPU as Platform service in the cloud through remote GPU sharing:

- **Cost**: GPUs are expensive. Having GPU installed on every machine in the cluster is infeasible and economically unsustainable. Moreover, GPUs are not a primary component of a computer system, consequently, CPUs are a primary component that is designated to run both system softwares like operating system along with the user applications. Contrary to GPU, this makes CPUs as an integral part of every machine.
- **Utilization**: GPU sharing in the cluster increases the resource utilization of the computing system.
- **Energy Efficiency**: GPUs are power hungry devices; efficient resource utilization saves a huge amount of energy that would have been spent unnecessarily.

Although the benefits of GPU remote sharing for cloud applications are manifold, it has lots of growing challenges and shortfalls to provide GPGPU as PaaS in the cloud.

*Challenges for GPU Programming in Cloud*

The following are the issues that inhibit the usage of GPUs in the cloud.

- **Non-transparent Remote Access**: GPU devices across a network provides an indirect non-transparent remote access. Resource requests must be wrapped with remote APIs and data should be serialized, marshaled and deserialized, this introduces performance overhead.

- **Virtualization**: Virtualization technology in cloud data center injects another level of complexity for GPU usage. Unlike CPU, GPU cannot be easily virtualized either by time sharing or space-sharing using existing hypervisor technologies. There are some recent efforts to virtualize some of the NVIDIA cards [8], however, most of them are for restricted to graphics rendering purpose and very few supports general-purpose computations.

- **Lack of Programming supports for Cloud-Native Applications:** The cloud-native application development process needs advanced tools and APIs to efficiently build the systems. However, lack of such tools for incorporating the GPUs hinders GPU usage in the cloud.

- **Lack of support for Windows platform:** Most of the customized frameworks that support GPU sharing are targeted for Linux platforms. There have been very few efforts for GPGPU regarding Windows environment where most of the today's desktop applications are built on top of the .NET framework.

- **Dynamic Resource Provisioning and Scheduling for GPGPUs.** Traditionally, clouds are used to adapt to the varying resource demand by dynamically provisioning and de-provisioning new resources at runtime. Despite GPU resource provisioning at the cloud,

> GPUs lack this critical future of dynamic provisioning and scheduling to benefit from adaptive resource usage that reduces huge amount cost.

The GPU remote sharing in the context of providing it as a platform service is the focus of this study. To achieve this, Aneka is extended which acts as middleware that provides integral resource management services.

## 4. Related Work

Using the GPUs in the cloud or virtualizing in a cluster has been a challenging work that many researchers have tried to solve in recent years. There are many approaches followed to virtualize or remotely access the GPU. Figure 2 shows the brief taxonomy of existing solutions. In this regard, the relevant works are described in this section.

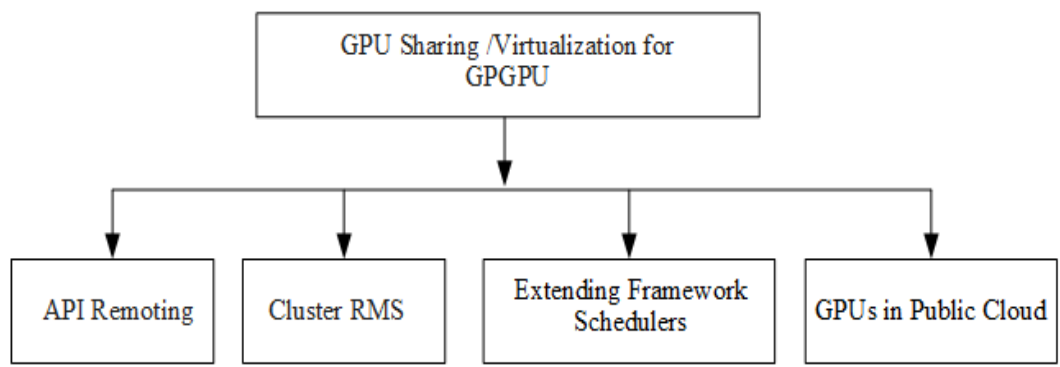

**Figure 2 Taxonomy of Shared GPGPU Programming**

### 4.1. GPU Virtualization through API Remoting

Several methods have been proposed to use GPUs in shared or virtualized environment. The most widely used technique is to intercept CUDA API calls at local machine and redirect it to a remote machine that has GPU access. The computation is done on a remote machine and the result is sent back to the requested machine.

**vCUDA**

vCUDA [9], elucidates GPUs access in a virtualized cloud environment for CUDA programming, it is built upon a client-server architecture. The CUDA API calls are bundled and marshaled with the necessary parameters and required information and further redirected to the remote host that has direct GPU access. It uses XML-based Remote Procedure Call (RPC) to transfer the intercepted call to host (remote) machine. A special stub is used at the host, which handles request from the guest OS and un-marshals the message and calls underlying CUDA device driver to execute.

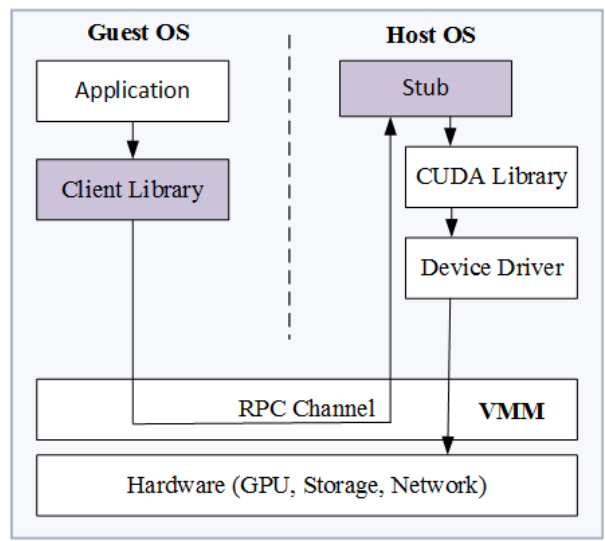

**Figure 3 The Schematic Overview of Remote Intercept Method** [9]

Figure 2 illustrates the generic architecture of the API remoting technique. In this architecture, every invocation to CUDA API from guest OS is intercepted and forwarded to the remote host machine. Note that, this method is useful for GPU sharing among virtual machines within a host or within the cluster.

**gVirtus**

gVirtus [10] is another library that relies on API remoting method. Like vCUDA, it also intercepts CUDA API calls, however, it utilizes efficient communication techniques by using communicators provided by different hypervisors that depend on TCP/IP protocol. gVirtus encompasses the virtualization layer and communicates directly to host-OS. Therefore, it is efficient but highly dependent on the communicators provided by hypervisors, like XENLoop, vmSocket, VMware Communicator Interface, etc.

*rCUDA*

rCUDA [4], remote CUDA is designed for GPU clusters. Initially built for clusters, rCUDA is later extended for Virtual Machines (VM). This framework offers GPU sharing solution for HPC clusters, a way of reducing the total number of GPUs in the cluster by utilizing the GPUs in remote machines, which are regarded as GPU servers in this framework. This framework mainly consists of two software components, a client module, and a GPU server module. It uses several optimization technologies like RDMA and it is by far the most robust GPU-sharing technology among the many existing solutions. In the case of VMs, client module software is installed on guest OS and server module is installed in the host OS.

These standalone libraries provide a way to perform GPU programming in virtualized or physical cluster. However, these solutions lack several capabilities of PaaS framework such as resource provisioning, monitoring, etc. In addition, the remote APIs wrappers must be written for every CUDA APIs and signatures of these APIs change rapidly, and new APIs and functionalities are included with their frequent new software updates, this makes developers to continuously update the APIs which is a cumbersome process.

### *4.2. Using Resource Management Systems*

The other prominent approach employed for distributed GPU programming or GPU sharing is extending the schedulers of popular cluster Resource Management Systems (RMSs). Some of the works have explored using the RMS like TORQUE [11], SLURM [12] and HTCondor [13]. In such workaround solutions, jobs can specify their request for the GPUs but it is left up to the user to ensure that the job is executed properly on the GPU [12]. Jobs are tagged to indicate GPU and non-GPU jobs and RMSs scheduling policies have extended to be aware of GPU nodes and the respective jobs are scheduled from the job queue. For example, Netflix has extended celery system, a distributed task queue system for their distributed machine learning application that was deployed on Amazon AWS with GPU instances. Nevertheless, this approach is application specific and mostly restricted for the HPC cluster setup and does not suit for generic application developments.

*4.3. Using Framework Schedulers*

Cluster programming languages like MPI and data processing frameworks like Spark and Hadoop have been extensively used for the distributed GPU programming. Coates et. al have demonstrated performing deep learning tasks using GPU enabled HPC systems. They have built system using Commodity Off-The-Shelf High-Performance Computing Technology: a cluster of GPU servers with InfiniBand interconnection. The application has been built using C++ with MVAPICH2 [14], an implementation of MPI. MVAPICH2 handles all the low-level communication using InfiniBand including the GPU support.

SparkGPU [15] is an extension of the in-memory MapReduce framework Spark for GPU. It transforms a generic data processing system and enables applications to exploit the GPU resources in the cluster. It addresses several challenges like reducing the internal and external communication data by efficient data format and optimal batching modes for GPU execution. It also provides task scheduling capacity among CPU and GPU.

Integration of GPU in Hadoop Cluster has been explored in GPU-in-Hadoop [16]. The default Hadoop scheduler is used to schedule the jobs on nodes of the cluster. The JCuda, JNI, Hadoop streaming, and Hadoop pipelines technology have been used to achieve this. However, this approach assumes all the nodes in the cluster has access to GPU. Consequently, this approach is not economically feasible. In a similar way, Xin et al. [17] demonstrated the implementation of Hadoop framework with the OpenCL.

The framework coupled extensions provide an alternative solution, however, the applications will be restricted to specific frameworks which are designed for a specific set of applications. Moreover, managing these platforms adds more complexity to application logic and are also not designed for cloud-native applications.

### *4.4.GPUs in Cloud*

The inherent virtualization characteristics of cloud introduce additional complexity in GPU access among VMs. The elementary problem is that GPUs are not designed to be shared or virtualized. Unlike CPU, a single GPU cannot be time shared or space shared using hypervisor softwares due to its technological constraints. However, a GPU can be dedicatedly attached to a VM using GPU passthrough technology [18], most of the hypervisor software support this passthrough technology. Hence, one can have multiple GPUs in a machine that are dedicated to different VMs or devise a virtualization independent technique to share a GPU among the virtual machines. However, GPU passthrough is limited to provide access to the single virtual machine and cannot be encompassed across multiple nodes.

Amazon Offers EC2 P3[3] instances that have up to eight NVIDIA Tesla V100 GPUs. These machines are powerful yet do not offer any mechanisms to share the GPU among multiple nodes. Similarly, Google offers GPU machines in their accelerated cloud computing arena, they have the flexibility to attach multiple types of GPUs to any type of instances and can be pre-empted based on the load to save the cost.

NVIDIA virtual GPU (vGPU) have made possible to virtualize the GPU among virtual machines to accelerate remote virtual desktop interface (VDI) performance. These are designed to accelerate video encoding, decoding and also other graphics rendering operations and do not support the GPGPU on most of the GPU cards [8]. For example, vGPU has CUDA support for only GRID M60-8Q, M6-8Q cards.

Considering the limitations of remote GPU access in the cloud, it is imperative that existing solutions are not designed to support GPU sharing for GPGPU on the cloud. Here, the focus is on providing PaaS support for GPGPU application development for cloud-native applications.

## 5. Methodology for Aneka GPU Computing

In this section, the methodology for integrating GPU computing paradigm in the Aneka is presented and also shown how to effectively share the GPUs in the cluster. The base entities that have extended and scheduling policies that are used for GPU sharing in the cloud are briefly discussed.

### 5.1. Methodology

The current Aneka PaaS framework is extended to support GPU programming in Aneka Cloud. To achieve this, firstly, the master node should be aware of the worker nodes that have direct GPU access (connected through the PCIe bus in the physical cluster, or through passthrough technology in the virtual cluster). Secondly, scheduling policy must be capable of mapping GPU tasks and nodes with the GPU access. Lastly, the GPU programming or tasks must be programmed with .NET supported programming languages.

---

[3] https://www.nvidia.com/en-us/data-center/gpu-cloud-computing/amazon-web-services/

As the default CUDA programming language does not have support for the .NET environment, there have been multiple efforts from the community to overcome this limitation and develop the coherent wrappers that provide simple APIs to program the GPU applications with .NET supported programming languages. These wrappers including CUDAfy[4], managedCUDA[5], Alea GPU[6] are some of those that are currently available to program the GPU-based applications in the .NET environment. In this chapter, CUDAfy and Alea GPU have been tested with the Aneka SDKs. The Alea GPU library is actively updated and supports a vast number of prevailing CUDA libraries like cuBLAS [19] and cuRAND [20], and it is also available on standard .NET package manager NuGet. Therefore, this library is used for the case study experiments.

Our proposed methodology aims to achieve the following objectives:

- **Programmability**: The integrated high-level APIs of Aneka SDKs along with .NET supported CUDA programming language manifests the efficient way of developing an application to utilize heterogeneous resources.

- **Ease of use:** The low-level communication, message (de)serialization and network aspects are automatically managed by integral management services of Aneka. Users can unreservedly focus on the actual business logic of the applications.

- **Cost Reduction:** The reduction in the number of GPUs in cluster benefits economically by reducing excess amount of resource acquisition and operational costs.

---

[4] https://www.codeproject.com/Articles/202792/Using-Cudafy-for-GPGPU-Programming-in-NET
[5] https://github.com/kunzmi/managedCuda
[6] http://www.aleagpu.com/release/3_0_4/doc/

## 5.2. Extended Aneka Architecture

The extended architecture of Aneka is shown in Figure 4. The new GPU-related components that are included in its architectural framework are highlighted. They include:

- Including GPU resources in IaaS
- Extending task programming model for GPU
- Providing easy to use SDK's for GPGPU on Aneka Cloud

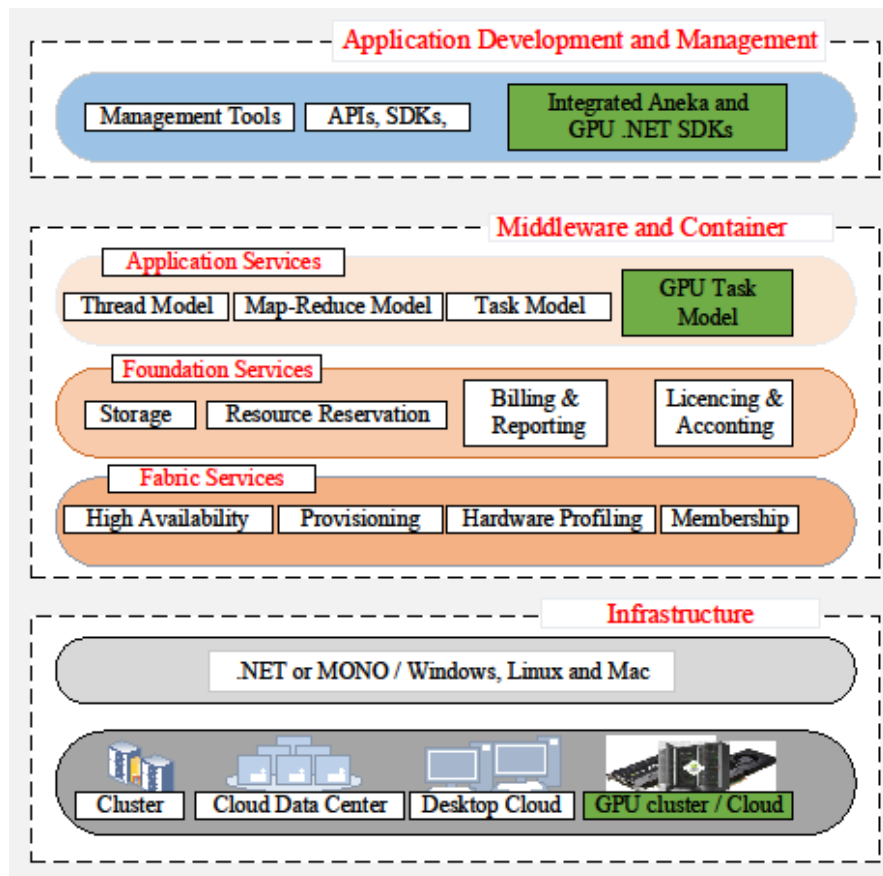

**Figure 4 Extended Aneka Architecture for GPU Programming**

### i. GPU Resources in IaaS

The present architecture of Aneka allows to develop and deploy applications which target CPU as a computing resource. Previously, at the infrastructure layer, all resources included were consisting of CPU computing resources only. In this architecture, GPU resources are also included in the IaaS. The computing infrastructure can be a standard physical cluster, virtualized cloud cluster or desktop-based private cluster. In this aspect, it is assumed that one or more nodes in the cluster have direct GPU access.

ii. *Task Programming Model for GPU*

The bag of tasks programming model is extended for GPGPU in which tasks are independent work units which can be executed or scheduled in any order. This feature can be effectively leveraged to execute the GPU tasks. In this regard, the current Aneka Task model is extended for GPU programming. Each CUDA kernel is wrapped as an independent task and it can be executed independently. This can be contemplated as a new application model from Aneka.

iii. *APIs and SDKs for GPGPU on Aneka Cloud*

At the application development layer, extended Aneka SDKs along with .NET wrappers for CUDA programming can be leveraged to develop the GPGPU applications. The required entities of the Aneka that are responsible to profile GPU enabled worker nodes and schedule accordingly is extended and built. The further explanations can be found in the next section.

5.3. *GPU aware Scheduling of Tasks*

The important feature of any cloud PaaS platform is to efficiently schedule the user tasks/jobs on cloud infrastructure in a user-transparent way. To this aspect, Aneka has separate Runtime scheduling service that transparently schedules user tasks on available resources. It is also responsible for interacting with several other services like active worker node membership catalog, dynamic resource provisioning services, heartbeat services, etc. The default task scheduling policy is to map tasks to the resources in First come First Serve (FCFS) in a round robin fashion. The resource capacity of each node is unified based on CPU capacity (MegaHertz) and tasks are assigned accordingly. The default scheduling interfaces in Aneka can be extended and new advanced policies can be incorporated to improve the performance and cost benefits based on application demand [21] [22].

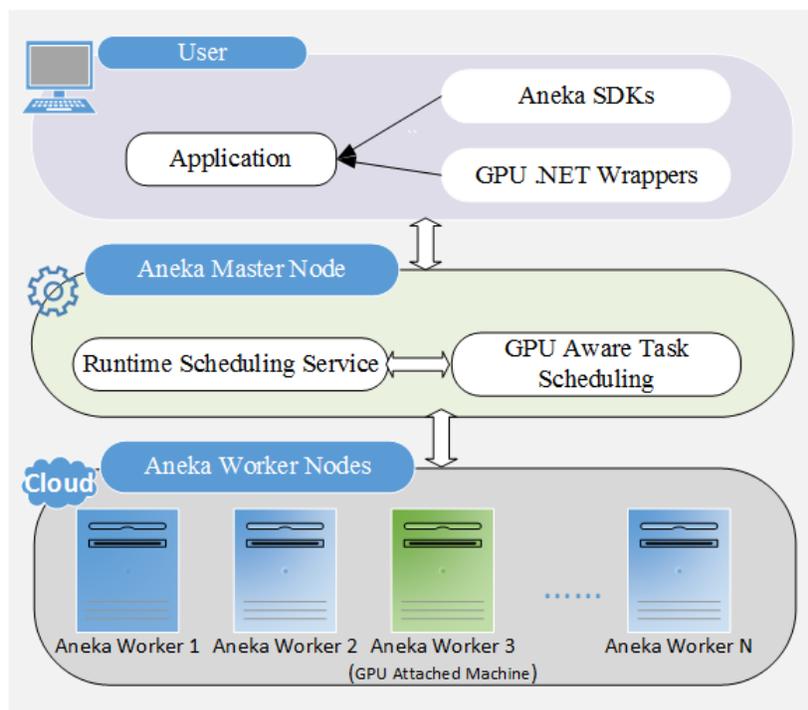

**Figure 5 The Overview of GPU-Aware Scheduling of tasks in Aneka**

A schematic overview GPU aware scheduling in Aneka is illustrated in Figure 5. It is assumed that the applications are constituted with multiple independent tasks where some tasks are CPU bound and some other are GPU bound tasks.

To accomplish the GPU aware scheduling, Aneka Runtime Scheduling service has been redesigned. Firstly, the tasks are differentiated into GPU tasks and non-GPU tasks, this is achieved by introducing a simple parameter into Aneka Task entity that tags an attribute to note GPU and non-GPU tasks, respectively. Secondly, the worker nodes are also segregated into two categories based on their ability to have direct GPU access. Finally, the Runtime Scheduling entity maps GPU tasks to only those nodes that have direct GPU access, and remaining CPU-based tasks are mapped to the rest of the machines based on CPU capabilities.

CPU bound tasks follow existing Aneka policy and GPU bound tasks are scheduled in a round robin fashion on GPU enabled nodes. In the present system, the capabilities of a GPU (number of cores and memory) and current load is not considered for scheduling decision. This can be considered in the future as a performance improvement.

### *5.4. Template for Aneka GPU Task Programming*

In this section, a sample template is outlined that shows how to write GPU program using extended Aneka task programming model. Aneka task programming model provides the flexibility to create a number of tasks and add it into the task pool. Once the task pool is submitted for execution, the master node distributes the tasks across worker nodes based on selected scheduling policies. By default, *Execute()* method of the task class is invoked that initiates the task execution. The following template of .NET class shows a sample schema for writing the Aneka GPU tasks.

*<summary>*

    // Every Aneka class implements interface called ITask, and this class is serializable

    // Public data members or public class members. The output of the GPU task must be written to these data members or shared file systems of Aneka containers.

    // Only these data members can be accessed by master node once the task is complete

*</summary>*

**public class AnekaGPUTask: ITask {**

    *<summary>*

        // constructor to initialize the data members

        <parameters, all the input data members or class members can be initialized by through these

      parameters>

*</summary>*

**public AnekaGPUTask (){**

    // initialize the data members of the object

**}**

*<summary>*

    //This is a start method and can be used to create a GPU handler and also for launching the kernel

    //<parameters, it can take any parameters and returns void always>

*</summary>*

**public void Execute(){**

    //Initialize the CUDAfy/ Alea GPU object or module

    // Set the GPU handler

    //Perform GPU memory allocation and copy the input data

    //Set the appropriate blocks and threads; this decides total number of threads we generate

    // Launch the kernel. or GPU Method Invocation

    //Copy back the result from GPU to public data members of Aneka Task class using APIs

**}**

*<summary>*

    // This method acts as a GPU method that executes on GPU; all the threads work in parallel in a SIMD manner.

    //<Parameters, all the input data, output data members or variables>

*</summary>*

**public void KernelMethod(){**

    //Map the thread id to the data index and define actual compute logic

    **}**

**}**

This template precisely presents how to write the GPU tasks in Aneka, other required necessary functionalities like how to create tasks, submit tasks, access the result once the tasks are finished and handling other events remains same as before. The only difference would be to set the flag *isGPUTask* to the true value which is part of extended Aneka Task object. For more detailed information about building and executing Aneka applications, readers are advised to refer the Aneka programming documents [23].

## 6. Image Edge Detection on Aneka Cloud using GPU: A Case Study

An image processing application built for edge detection of objects in an image has been chosen to evaluate the feasibility and performance of modified Aneka system. Both sequential and parallel algorithm and its GPU version of Aneka have been implemented. The experiments were carried out on a private desktop cluster. The implementation details, results and outcomes are discussed below.

### *6.1.  Edge Detection Algorithm*

Detecting edges of an object in a digital image are one of the essential applications of image processing.  It requires a high amount of computational resources and time. There exist several methods to perform this operation using different operators or filters. here, the commonly adopted Sobel operator [24] is used.

To find the edges of objects in an image, one needs to compute the new value for every pixel based on the certain operation. In the Sobel filter, the output value of pixels is calculated based on the Equation 1. The value of $S_x$ and $S_y$ in the Equation 1 can be derived based on the Equation 2.

$$P = \sqrt{S^2_x} + \sqrt{S^2_y} \qquad (1)$$

$$S_x = \begin{bmatrix} -1 & -2 & -1 \\ 0 & 0 & 0 \\ 1 & 2 & 1 \end{bmatrix} * M_i \quad S_y = \begin{bmatrix} 1 & 0 & -1 \\ 2 & 0 & -2 \\ 1 & 0 & -1 \end{bmatrix} * M_i \qquad (2)$$

In Equation 2, $S_x$ and $S_y$ are resultant value produced by the product of 3×3 filter matrices with $M_i$; where the matrices represent horizontal, and derivative approximations. $M_i$ is a 3×3 matrix for pixel i that is constituted by including its all eight neighbor pixels from the matrix, where i = 1 to m*n, m is the width of the image and n is height.

It is important to note that finding the value of P for each pixel is independent of other pixels. Hence, this embarrassingly parallel property can be exploited using GPU architecture which supports such inherent data-parallel applications.

In sequential computation on CPU, all the pixels values are calculated in a sequential manner. However, this increases the execution time of applications because the complex image may have millions of pixels. To reduce this computational time, this application is adapted for GPU architecture which runs faster than the only CPU versions by using abundant GPU cores and running parallel with GPU threads.

### 6.2. *Parallel Implementation of Sobel Edge Detection on GPU's*

In the parallel implementation of edge detection application, the value of each pixel is calculated in parallel with a dedicated sperate GPU thread. At first, the image matrix is converted into the byte array, then, the byte array is transformed into integer array to perform the matrix operations.

The CUDA kernel takes the array of pixels as input and returns the output array. Each pixel value is calculated by separate CUDA thread where all the threads run in parallel. The output pixels are written into a separate output array. This output array is later copied to CPU memory using the CUDA APIs.

In the GPU version implementation, K (K= image width * image height) number of CUDA threads are generated. Inside CUDA kernel, each thread index is mapped to corresponding data (pixel) index based on the thread id and pixel number.

Parallel version uses Aneka Cloud where its scheduling service is responsible to schedule the GPU tasks on appropriate nodes. The image read, deserialization (converting to a byte array), serialization (converting output byte array into image matrix) is done by Master thread which runs on Master node and edge detection operation is wrapped into Aneka GPU task that runs on a GPU equipped worker node.

In Aneka GPU implementation, the *Execute()* method of Aneka task class, which is the default method to begin task execution is responsible for querying the GPU device and setting up the handler for the GPU device. Further, another method is invoked with the GPU handler to perform the main application logic. The output image is written to the Aneka container file system, alternatively, if the shared file repository has been set up, the output can be stored in the repository from where all nodes in the cloud can access the data.

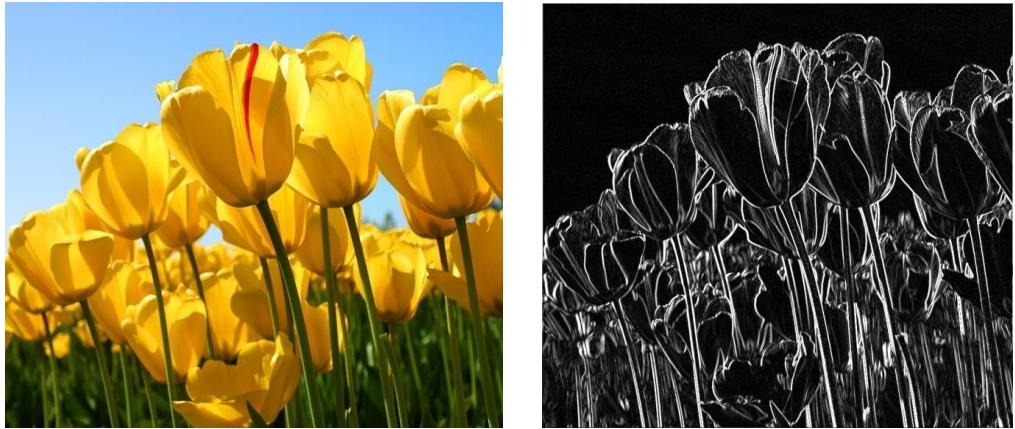

Input Image                        Output Image

**Figure 6: Edge Detection Sample Input and Output Images.**

### *6.3.  Results and Analysis*

The experiments are conducted on a 4-node desktop-based cluster. Aneka PaaS framework has been installed on all the four nodes. One node is designated as Aneka Master and three nodes were assigned worker roles. All the nodes have preinstalled with Windows 7 as their operating system. In this setup, one machine from worker nodes is equipped with NVIDIA QUADRO K620 GPU Card (consumer grade card). This GPU has 384 computing cores and 2 GB of GPU memory, CUDA driver 9.1 is installed on this machine.  The application has been built using C# console on .NET framework 4.0 with Visual Studio 2015.

| Image Size | Sequential (time in ms) | Aneka GPU Task (time in ms) |
|---|---|---|
| $512 \times 512$ | 60 | 75 |
| $1024 \times 586$ | 185 | 115 |
| $5184 \times 3456$ | 5437 | 887 |

**Table 1.  Results of Sobel Edge Detection Algorithm.**

The access to the input image is given to the worker node directly from the local storage. In a real public cloud setup, data is stored in some shared storage servers and these storage servers are usually mounted on all the worker machines or have high-speed network access. The local access to input image overcomes large overhead that would have been introduced by network transfer of the input data.

To evaluate the performance of the implementation, different dimensions of images are considered. The following sample image shown in Figure 7 has a dimension of 1024×768 (width and height). Here, the CUDA kernel is launched with 1024×768 threads. All the threads calculate resultant pixel value using formula explained in Equation 1. To analyze the effect on execution time, several other images with different dimension have been used. Table 2 shows the detailed timing analysis with a different dimension of images.

As shown in Table 2, for the larger images, GPU implementation performs better. The timing includes local GPU computation time at worker nodes. The Aneka framework has overhead which includes all the network serialization and deserialization of code and other necessary functionalities. It is almost a constant (around 2000 ms, empirically calculated) time as it is independent of input size and the size of the code remains the same. So, to get the clear performance benefit, our input should be large enough to absorb this overhead. In this experiment, it is only possible to consider maximum dimension image of $5184 \times 3456$. This is because of the limited memory on GPU (2GB) in our consumer grade GPU card. If the experiments are run with enterprise level GPUs such as NVIDIA Tesla [25] cards and with images of higher dimension which is a usual case in today's digital platforms (for example, video rendering/ streaming or online gaming), performance benefit will be clearly visible. This evaluation can be considered in the future.

To conclude, this case study demonstrates how to develop cloud-native GPU applications with remote GPU sharing. The results have demonstrated performance benefit and feasibility of Aneka platform with GPU computing model.

## 7. Future Directions

The elementary principal of cloud computing is elasticity to provide resource provisioning based on application demand to reduce the cost and meet Service Level Agreements (SLA). Towards this, Aneka can be extended to dynamically provision and deprovision the GPU instances on public cloud based on the application demands. This can be achieved by extending Aneka provisioning service and utilizing the public cloud REST APIs and further considering application-specific profiles and load. For example, the deadline-driven provisioning for scientific application for Aneka is studied in [22]. In a similar way, application specific dynamic provisioning of integrated CPU and GPU based instances can be investigated.

Another useful avenue to explore is fine-grained resource capacity aware scheduling policies. The GPU capabilities differ from different types of cards. The scheduling policy can consider the number of cores, the memory capacity of GPU instance for scheduling decision.  In a similar way, another key area to evaluate is the parameter sweep model with existing GPU applications. Instead of redeveloping GPU applications with Aneka SDKs and GPU wrappers, the executables can be used with different input data range to directly execute on Aneka cloud. In a similar line, GPU computing support can be extended to thread and map reduce model-based applications.

## 8. Summary and Conclusions

The GPU computing is gaining popularity for executing many scientific and business applications. In addition, cloud computing has become a mainstream paradigm to deliver subscription-oriented computing services. Consequently, having GPU on every machine in a cluster is expensive and infeasible. Therefore, it is important to have useful tools and frameworks to easily develop and deploy the applications and share the underlying resources effectively to reduce the cost.

In this chapter, the GPGPU PaaS model is presented with Aneka and also identified the challenges in realizing the GPU sharing in a cloud scenario. The relevant works have been studied and categorized them into a taxonomy. The methodology adopted for integrating GPGPU model into Aneka framework is described. The feasibility of the proposed solution is demonstrated through a case study with an image processing application that performs edge detection of objects in an image. This chapter also identifies important future directions in this domain.

The support for GPUs as part of Aneka PaaS environments enables rapid creation and deployment of big data applications on clouds at the same time harnessing GPU capabilities for performance and energy efficient computing.